# Particle size dependence of orbital order-disorder transition in LaMnO$_3$

Nandini Das, Parthasarathi Mondal, and Dipten Bhattacharya[*]
*Electroceramics Division, Central Glass and Ceramic Research Institute,
Calcutta 700 032, India*

The latent heat ($L$) associated with the orbital order-disorder transition at $T_{JT}$ is found to depend significantly on the average particle size ($d$) of LaMnO$_3$. It rises slowly with the decrease in $d$ down to ~100 nm and then jumps by more than an order of magnitude in between $d$ ~100 nm and ~30 nm. Finally, $L$ falls sharply to zero at a critical particle size $d_c \approx 19$ nm. The transition temperature $T_{JT}$ too, exhibits almost similar trend of variation with the particle size, near $d$~30 nm and below, even though the extent of variation is relatively small. The zero-field-cooled (ZFC) and field-cooled (FC) magnetization vs. temperature study over a temperature range 10-300 K reveals that the antiferromagnetic transition temperature $T_N$ decreases with $d$ while the temperature range, over which the ZFC and FC data diverge, increases with the drop in $d$. The FC magnetization also, is found to increase sharply with the drop in particle size. A conjecture of non-monotonic variation in orbital domain structure with decrease in particle size – from smaller domains with large number of boundaries to larger domains with small number of boundaries due to lesser lattice defects and, finally, down to even finer domain structures with higher degree of metastability – along with increase in surface area in core-shell structure, could possibly rationalize the observed $L$ vs. $d$ and $T_{JT}$ vs. $d$ patterns. Transmission electron microscopy (TEM) data provide evidence for presence of core-shell structure as well as for increase in lattice defects in finer particles.

The orbital physics assumes significance in strongly correlated electron systems[1] like perovskite manganites or high-$T_c$ superconductors not only because understanding the orbital physics provides vital clues to novel phenomena like metal-insulator transition, superconductivity, colossal magnetoresistance, phase separation, interplay among orders in charge, spin, orbital degrees of freedom etc. but also because of possible future applications of orbital electronics[2] in many areas of micro-electronics. Recent works have shown how different orbital phases – solid, liquid, glass, liquid crystal etc. – develop across a whole range of doped and undoped manganites.[3] Study of orbital order-disorder transition from both thermodynamics and kinetics points of view helps in ascertaining the nature of the orbital phases. Local structural studies namely, resonant X-ray scattering (RXS)[4] or coherent x-ray beam scattering[5] provide information about the orbital domain structure and its variation as a function of overall lattice distortion and charge carrier doping. It seems that through all these studies a comprehensive picture of orbital phases and their role in governing a range of novel phenomena observed in strongly correlated electron systems is emerging.

It is interesting, in this context, to study the orbital order-disorder transition in confined geometries, e.g., in nano-structured systems. Nano-structured systems offer novel properties like improved magnetization, quantum effects, improved mechanical property etc.[6] By proper tuning of the particle size, it is possible to optimize the desired property – electrical, magnetic, optical, thermal, mechanical etc. The phase transition characteristics too, undergo drastic changes with the decrease in particle size down to less than 100 nm.[7] Sophisticated techniques like ultra-sensitive nano-calorimetry,[8] scanning tunneling microscopy together with perturbed angular correlation,[9] laser irradiation coupled with local calorimetry[10] etc. have been used in order to study the melting phenomena in nano-particles of metals. It has been observed that both the melting point ($T_m$) as well as the latent heat ($L$) depress as inverse of atom cluster radius ($r$) down to a cluster size of ~2 nm.[8-11] The phenomenon of melting follows either homogeneous melting[12] or liquid-shell melting[13] or nucleation and growth kinetics model.[14] The depression in $T_m$ or $L$ results from increased surface area where atoms are rather loosely bound.

The study of orbital order-disorder transition in nano-structured $LaMnO_3$ assumes significance in the context of use of such nano-structured systems in micro-electronics applications. It is also important in the context of understanding the variation in orbital domain structure as a function of particle size and thereby variation in spin structure. In this paper, we report results of orbital order-disorder transition studied by conventional global calorimetry in a series of nano-structured $LaMnO_3$ systems with different particle sizes. The latent heat ($L$) associated with the transition is found to follow a non-monotonic pattern with the particle size $d$. The transition temperature ($T_{JT}$) is also found to follow non-monotonic pattern. While the rise in $L$ is more than an order of magnitude near $d \sim 30$ nm, the $T_{JT}$ is found to have first dropped by more than ~10% with the drop in particle size and then exhibits a rise by a few degrees near $d$ ~30 nm before finally decreases by nearly ~30% with further decrease in particle size down to the critical size $d_c \approx 19$ nm. The latent heat ($L$) also drops precipitously within this particle size range. To the best of our knowledge such results are being reported for the first time. In order to rationalize the observation we conjecture a variation in orbital domain structure as a function of particle size together with increase in surface area in core-shell structure of particles. The TEM results provide information regarding core-shell structure and lattice defects as a function of particle size.

The nano-crystalline LaMnO$_3$ particles have been prepared by various techniques. For this work we have primarily resorted to bottom up approaches[15] like preparation of fine particles by using micro-emulsion[16] or precipitation from a solution under ultrasonic vibration.[17] These techniques have produced phase pure fine particles. In the case of micro-emulsion technique, a solution is prepared by mixing aqueous solutions of La(NO$_3$)$_3$ and Mn-acetate in 1:1 concentration ratio along with cetyltrimethyl-ammonium bromide (CTAB) as surfactant, n-butanol, and n-octane. The aqueous metal nitrate/acetate solution together with CTAB forms the micro-emulsion within n-octane. The n-butanol is added as a co-surfactant. Another emulsion having oxalic acid, CTAB, n-butanol, and n-octane is prepared. In this case, aqueous solution of oxalic acid together with CTAB forms the micro-emulsion within n-octane. This emulsion is added to the previous one under constant stirring. Each emulsion droplet acts as a nano-reactor. The emulsion droplets from the added solution coalesce with those in the earlier one and facilitate the precipitation reaction within the droplets whose size is restricted by the surfactant. The size of the droplets can be varied by changing the concentration of the surfactant which, in turn, gives rise to particles of different sizes. In the case of micro-emulsion technique, the concentration of the surfactant is increased from 8.5 to 9 millimol/litre for reducing the particle size down to 20 nm. The precipitate, thus obtained from the resultant mixture, was centrifuged and washed with acetone and alcohol mixture. Finally, the powder was dried under vacuum at ~50°C. In the case of sonochemical technique, aqueous solutions of lanthanum nitrate and manganese acetate were mixed and sonicated by using Ti horn (20 kHz, 1500 W, Vibracell, USA) for 2h at room temperature. Small amount of decalin is added for efficient power transfer. To reduce agglomeration, SDS surfactant was added under sonication. Here too, the concentration of the surfactant is varied from 0.85 to 1 millimol/litre for controlling the particle size. Oxalic acid is also added during sonication. Sonication is continued till precipitation is complete. Finally, the precipitate was collected, washed with alcohol and acetone and dried under vacuum at ~40°C. The powders prepared by both the techniques were subject to heat treatment under different temperatures for different time span. The temperature was varied over 700-800°C while the time was varied over 4-6h. The controlled variation in surfactant concentration as well as heat treatment time and temperature helps in preparing particles of different sizes. These parameters have been optimized in order to prepare powders with requisite particle size. The X-ray diffraction (XRD) patterns exhibit crystalline perovskite orthorhombic phase with Pbnm space group. A representative XRD pattern is shown in Fig. 1(a). In Fig. 1(b) we show, representative high angle X-ray lines. From such line broadening one can normally calculate the average crystallite size ($d$) by using Debye-Scherrer model. The Debye-Scherrer model can be written as $d = 0.9\lambda/\beta_d cos\theta$, where $\lambda = 1.54056$ Å for Cukα line, $\beta_d$ is the full width at half maximum of an x-ray diffraction peak, and $\theta$ is the corresponding Bragg angle. The lattice strain ($\varepsilon$), on the other hand, as a function of particle size can be calculated using the Wilson formula $\varepsilon = \beta_\varepsilon/4tan\theta$ where $\beta_\varepsilon$ describes the structural broadening, which is the difference in integral X-ray peak profile width between the sample and a standard (silicon) and is given by $\beta_\varepsilon = \sqrt{\beta_{\varepsilon,obs}^2 - \beta_{std}^2}$. Silicon standard is used as it has a large crystallite size. However, since the actual X-ray peak broadening in the present case is taking place due to both the decrease in particle size and increase in lattice strain, we followed the Williamson-Hall approach to calculate the particle size ($d$) and the lattice strain ($\varepsilon$). In this approach, the experimental X-ray peak broadening ($\beta_{exp}$) is given by $\beta_{exp} = \frac{0.9\lambda}{d \times \cos\theta} + \frac{4\varepsilon \sin\theta}{\cos\theta}$. By plotting $\beta_{exp}.\cos\theta$ as a function of $4\sin\theta$ for several X-ray diffraction peaks, we obtained a straight line. This linearity shows that for our case, the Williamson-Hall approach is applicable. The slope of the straight line gives the lattice strain ($\varepsilon$) while the crystallite size ($d$) is calculated from the intercept of the straight line with the vertical axis. The lattice strain is found to be compressive and varying between 0.22%-0.48% with higher strain for finer particles [Fig. 1(c)]. The lattice strain appears to have increased at a faster rate below $d$ ~30 nm. In Fig. 2, we show the representative TEM photographs of the nano-particle assemblies. Using image analyzer, we determined the average particle size as well as the histogram of the particle size distribution. The average particle size is found to vary between 20-100 nm (aspect ratio 1.25-1.27). While presenting our data on nanoscale $LaMnO_3$, we use the average particle size estimated from the TEM study. We notice a bit of difference between the crystallite size estimated from XRD peak profile analysis and the particle size estimated from TEM study. This difference is due to the presence of lattice strain which influences the XRD peak profile. The core-shell structure of the particles can be clearly observed in TEM photographs of the finer systems. The high resolution TEM (HRTEM) picture helps in finding out the average lattice plane spacing. It varies between 0.2-0.4 nm. One interesting observation is the variation in lattice defect structure as a function of particle size. In finer particles, one observes defect lines and different facets in lattice structure [Fig. 2(b)] while in relatively coarser particles one observes rather defect-free lattice [Fig. 2(a)]. We have measured the BET specific surface area of the powder as well. It is found to very between 16-20 $m^2/gm$. Poor surface area possibly results from agglomeration of the nano-particles. All these data are consistent and, therefore, the quality as well as the average particle size of the powders appears to be well defined. The $Mn^{4+}$ concentration is measured by chemical analysis (redox titration) and is found to be varying between 2-5%.

We have studied the orbital order-disorder transition by global calorimetry. In Fig. 3, we show differential scanning calorimetric (DSC) and differential thermal analysis (DTA) thermograms for a few samples having different particle sizes. Quite obviously, the phase transition peak position and area vary with the particle size. We have evaluated the latent heat by subtracting the background and calculating the peak area. We have also studied the low-field magnetic properties of the samples. The zero-field cooled (ZFC) and field-cooled (FC) magnetization vs. temperature measurements have been carried out across a temperature range 10-300 K under a field ~100 Oe. The variation in magnetization with temperature for a few representative samples is shown in Fig. 4. It is found that the antiferromagnetic transition temperature ($T_N$) decreases systematically with the particle size $d$. On the other hand, the temperature range over which the ZFC and FC data diverge ($\Delta T$) appears to increase with the decrease in particle size. For instance, for ~19 nm particle size, $\Delta T$ is ~255 K whereas for ~54 nm it is ~215 K. This implies increase in metastability. The FC magnetization increases with the decrease in particle size, which could be due to increase in surface magnetization. The variation of latent heat ($L$), $T_{JT}$, and $T_N$ with the average particle size is shown in Fig. 5. The calorimetric study has been repeated several times on a particular sample as well as on freshly prepared samples in order to verify the reproducibility of the data. In all these measurements, less than 10% variation in $L$ and less than ~2% variation in $T_{JT}$ could be observed. The heating and cooling cycle data depict small amount of "thermal history" effect in peak position. The peak position differs in heating and cooling cycle by less than 2 K. *These are the central results of this paper*. For the sake of comparison, we have included the data corresponding to bulk polycrystalline as well as single crystal $LaMnO_3$ samples. Good quality single crystal has been prepared by traveling solvent zone technique.

It is interesting to note that $L$ is varying significantly with $d$. It rises by more than an order of magnitude with the decrease in $d$ down to ~30 nm and then drops sharply to zero at a critical particle size $d_c$ ~19 nm. This pattern apparently resembles the one of coercivity ($H_c$) vs. $d$ observed in nano-crystalline ferromagnetic systems.[18] The transition temperature $T_{JT}$ too, exhibits similar pattern, even though, the rise is rather small. Such non-monotonicity in $L$ vs $d$ and $T_{JT}$ vs $d$ patterns is a significant departure from those commonly observed in melting in nanoscale.[8-11] This result shows that one can improve the enthalpy change or latent heat in $LaMnO_3$ around the orbital order-disorder transition point by resorting to nanoscale system. It also provides the limit of the particle size within which one is expected to observe such improvement. Such improved latent heat, if observed under electric or magnetic field or under pressure, can be utilized in different sensor applications.

It may also be noted, in this context, that normally both melting point as well as latent heat of melting are found to decrease systematically with $d$.[11] Here, even though, the latent heat of orbital order-disorder transition is found to rise by more than an order of magnitude, the rise in transition temperature $T_{JT}$ is small and is certainly not higher than that observed in bulk or single crystal sample. Therefore, it appears that rise in latent heat of transition is not always associated with concomitant rise in transition temperature. The patterns of $L$ vs. $d$ and $T_{JT}$ vs. $d$ can be understood, at least, *qualitatively* by resorting to the concept of orbital domain structure. It seems that the overall orbital domain volume determines the latent heat. In a continuum, the latent heat would have followed a monotonic pattern of decrease with the decrease in particle size. The *non-monotonicity* observed in the present case points out relevance of the discrete domains within a particle. The particle size alone does not determine the overall latent heat of transition at $T_{JT}$. In bulk LaMnO$_3$, the particles (or grains) are multi-domain in nature. They contain both domain and domain boundaries. The domain boundaries are disordered and, therefore, do not undergo a first-order transition at $T_{JT}$. The net orbital domain volume is relatively small in bulk systems. As the particle size decreases, we reach essentially a cleaner, i.e., defect-free system which, in turn, leads to a sharp rise in domain volume with very less domain boundaries. Of course, a direct observation of variation in orbital domain size as a function of particle size would have supported or discarded this conjecture. This experiment is beyond the scope of the present work. In the absence of such data we conjecture about the variation in orbital domain size as a function of particle size using the existing models[19] of grain growth during grain coarsening/sintering in a solid. The bigger grains grow at the expense of smaller grains during grain growth by two processes: (i) mass transport across the grain boundary and (ii) grain boundary motion. The grain growth can be slower if the radius ratio ($R$) between adjacent grains and the overall density of the matrix are below certain critical values.[20] In such a scenario, the grain growth takes place via the mass transport process. The growth can accelerate once these parameters cross the critical values. In that case, the process of grain boundary motion takes over. The anomalous $L$ vs. $d$ data, in the present case, can possibly be rationalized by invoking this grain growth scenario. The decrease in particle size could be considered as analogous to sintering/coarsening under pressure. There are, in fact, three different regimes across the entire range of $L$ vs. $d$ pattern – between ~100 nm to bulk (I), between ~29 to ~100 nm (II), and finally between ~19 to ~29 nm (III). In the regime-I, all the orbital domains could be of nearly equal size. They are small with sizable volume fraction of domain boundaries. Therefore, the overall latent heat ($L$) is small. The growth process is slow as radius ratio between adjacent domains ($R$) as well as the density of the matrix is low. This process continues till the particle size reaches ~100 nm. In the regime-II, the growth process accelerates as $R$ and the overall density of the matrix reach the critical values. The domain growth and sintering eventually lead to the formation of single domain structure near $d \sim 29$ nm. Similar scenario can be observed in packing of granular objects where percolation of force chains gives rise to sharp rise in packing density.[21] Such clean lattice is normally observed in thin films where one observes sharp improvement in properties compared to those of the bulk system. With further decrease in particle size in regime-III, the domain size becomes smaller. Therefore, the latent heat of transition also approaches zero in such a scenario. Such a decrease in domain size could result from two factors: (i) increase in lattice defects and disorder in finer particles and (ii) increase in lattice strain. In fact, over the entire regime of particle sizes, there is always a competition between homogeneous cleaner lattice and strained defective lattice volume fractions. It has been observed from X-ray line profile analysis that lattice strain does increase with the decrease in particle size. So, while the bulk particle contains sizable lattice defects and hence large domain boundaries with smaller domain volume, the particles become essentially defect-free and single-domain ones near an optimum size $d \sim 30$ nm. Even finer particles, of course, contain large defects, twins, as well as disorder, which, in turn, give rise to smaller domains. Such variation in lattice properties could be intrinsic size effect as powders prepared by two different techniques – microemulsion and sonochemical – result in nearly identical results. The plot of $L$ vs orbital domain size, across the bulk to nanoscale particles, is expected to exhibit a monotonically decreasing pattern. This pattern apparently corroborates the results obtained in a series of $La_{1-x}R_xMnO_3$ systems (R = Pr, Nd, Sm, Gd etc.) where $L$ decreases monotonically with average A-site radius $<r_A>$ and thereby with orbital domain size.[22] It is important to point out that in regime-III, one does not observe semblance of superpara-orbital ordered state analogous to superparamagnetic state with no finite $T_{JT}$ and $L$. Instead, both $T_{JT}$ and $L$ decrease systematically with the decrease in $d$ without dropping to zero abruptly. Direct experimental determination of orbital domain size as a function of particle size will clear the entire scenario and this will be attempted in a future work.

The $T_{JT}$ exhibits a drop with the drop in $d$. This is primarily because $T_{JT}$ does not depend much on the orbital domain size. Instead, it depends mostly on the average Mn-O-Mn bond angle $<cos^2\phi>$.[23] In thinner films, it has been found that $T_{JT}$ drops by ~10-20% due to increased lattice strain.[24] $T_{JT}$ also depends on the lattice disorder. Since in nano-particles, surface area increases and, thereby, gives rise to lattice disorder, $T_{JT}$ is found to drop systematically with $d$. The rise in $T_{JT}$ by a few degrees near $d \approx 30$ nm could result from a competition between increase in surface area and simultaneous decrease in domain boundaries in essentially defect-free bulk regions in a cleaner system. In smaller particles, disorder becomes more as surface area increases rapidly with the drop in particle size. Therefore, $T_{JT}$ is found to decrease again in smaller particle systems. Impact of domain boundaries on $T_N$ is not so prominent and, therefore, $T_N$ depicts no such non-monotonicity. It decreases monotonically with the decrease in $d$ which is quite similar to the decrease in $T_N$, observed in $RMnO_3$ with smaller R-ions like Pr, Nd, Sm etc due to chemical pressure effect.[23]

The HRTEM data shown in Fig. 2, provide evidence of increased lattice defects in finer particles and seem to support our conjecture of variation in orbital domain size with the decrease in particle size. For example, for the particles of average size ~ 32 nm, the lattice planes (fringes in the figure), within an isolated particle, are found to be oriented parallel to each other and free from domain boundaries or defects like twinning planes (Fig. 2a). The average lattice plane spacing is ~2.25 Å which corresponds to the (202) plane. The crystallographic axes are shown corresponding to the HRTEM image. On the other hand, in even finer particle system, presence of such defect structures within a particle is quite evident. The orientation of the lattice plane changes along defect lines within a small spatial scale. The average lattice spacing for one such set of lattice fringes is ~3.75 Å which corresponds to the (110) plane. Because of enhanced strain and lattice defects the orbital domain size is expected to be small in finer particles.

In summary, we observe that the latent heat ($L$) associated with the orbital order-disorder transition in pure LaMnO$_3$ varies significantly with the average particle size ($d$). One important observation is almost an order of magnitude rise in $L$ for a particle size d ~30 nm. Such a large $L$ can be exploited in many applications, if orbital order-disorder transition or melting of orbital order can be driven by photo-irradiation or current/magnetic field pulses. The non-monotonic pattern of variation of $L$ with $d$ could possibly be resulting from non-monotonic variation of orbital domain size as a function of particle size. Such variation in orbital domain size is possibly due to variation in lattice defects, strain, and disorder. The TEM, HRTEM data provide evidence of increased lattice defects and shell area in core-shell morphology of the particles with the decrease in particle size.

We thank S. Bose of Materials Research Center, Indian Institute of Science, Bangalore for carrying out the low temperature magnetic measurements and J. Ghosh of CG&CRI for helping in analyzing the X-ray diffraction data. We also thank P. Mandal of Saha Institute of Nuclear Physics, Calcutta for providing the single crystal of LaMnO$_3$.

**Figure Captions**

Fig. 1. (a) Typical room temperature X-ray diffraction pattern of the nano-particles of LaMnO$_3$. The average crystallite size corresponding to this pattern is ~25 nm. (b) Representative high angle X-ray peaks blown up. The full width at half maximum (FWHM) is used for calculation of the crystallite size. (c) The lattice strain vs. particle size (*d*) plot. The lattice strain appears to have increased at a faster rate below *d* ~ 30 nm. The solid lines are guides to the eye.

Fig. 2. (a) TEM and HRTEM photographs of the assembly of particles and lattice fringes within an isolated particle, respectively. The average particle size is found to be ~32 nm. The lattice planes appear to be defect free over a large region. The core-shell structure of the particles can be observed with shell thickness estimated to be ~ 5 nm. Arrow points to the shell region. (b). TEM and HRTEM photographs of the assembly of nanoscale particles and lattice fringes within an isolated particle, respectively. The average particle size is estimated to be ~20 nm. The lattice fringes appear to have changed the orientation across the defect line. The defect lines are evident and marked by arrows. The shell thickness is large and is estimated to be ~ 8 nm. Arrows point to a defect line.

Fig. 3. DSC (top) and DTA (bottom) thermograms for few of the LaMnO$_3$ systems with different particle sizes are shown. Top inset: DSC thermogram for the LaMnO$_3$ single crystal is shown. The rate of heating was 10$^o$C/min.

Fig. 4. Zero-field cooled (ZFC) and field cooled (FC) magnetization vs. temperature patterns for few of the nano-crystalline LaMnO$_3$ samples. The measurement has been carried out under an applied magnetic field 100 Oe.

Fig. 5. The latent heat associated with orbital order-disorder transition vs. average particle size in pure LaMnO$_3$. The variation of $T_{JT}$ and $T_N$ with the average particle size is also shown. The data corresponding to ~1 μm particle size represent those of bulk sample and single crystal. The data corresponding to the single crystal are marked by open up-triangle symbol ($T_{JT}$ ~ 740 K, $L$ ~ 30 J/g).

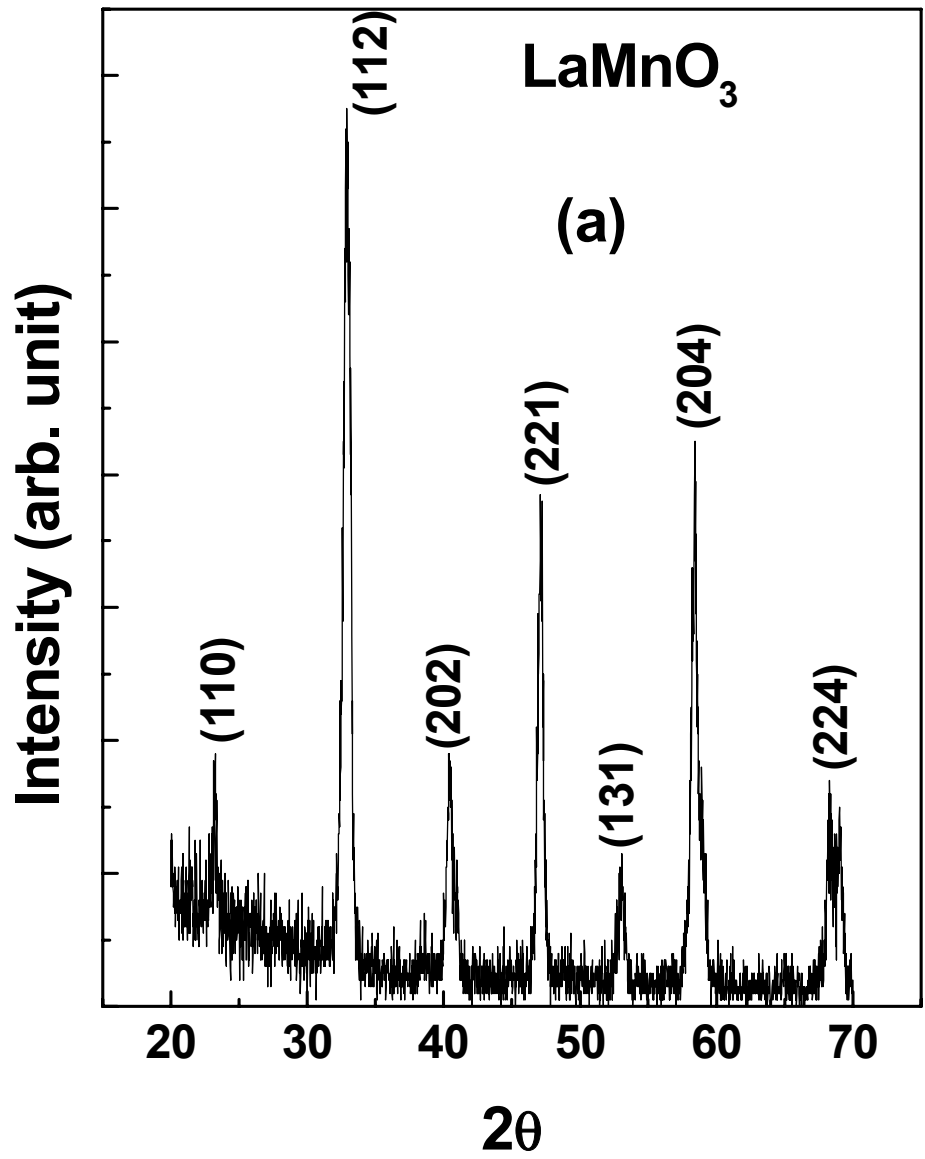

Fig. 1a

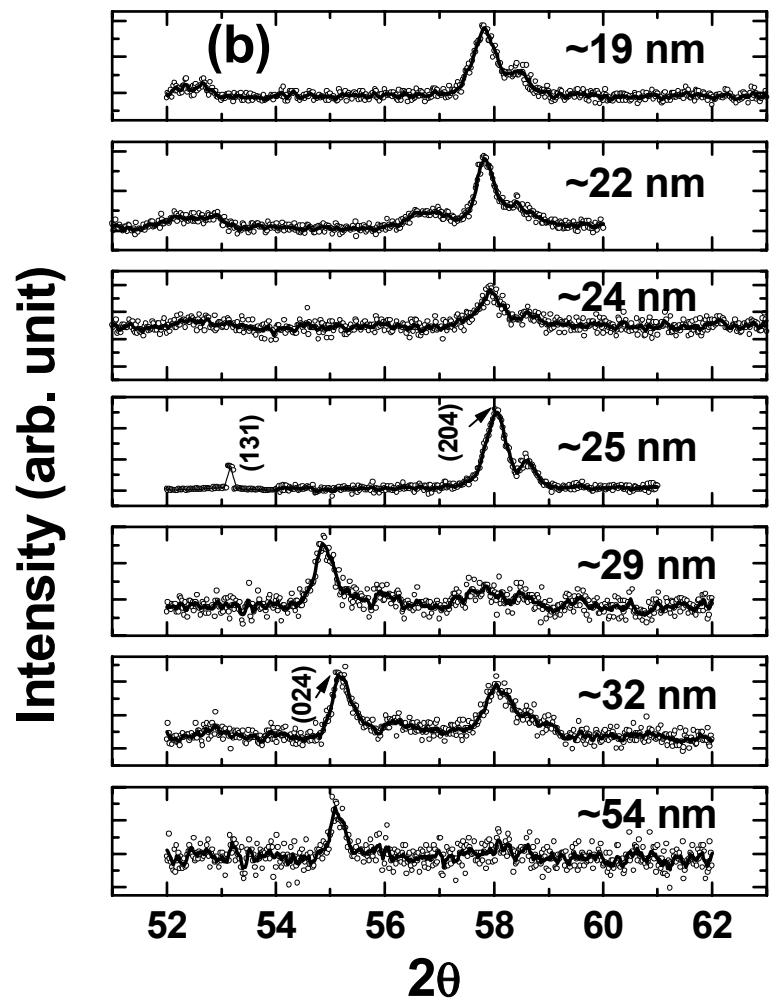

Fig. 1b

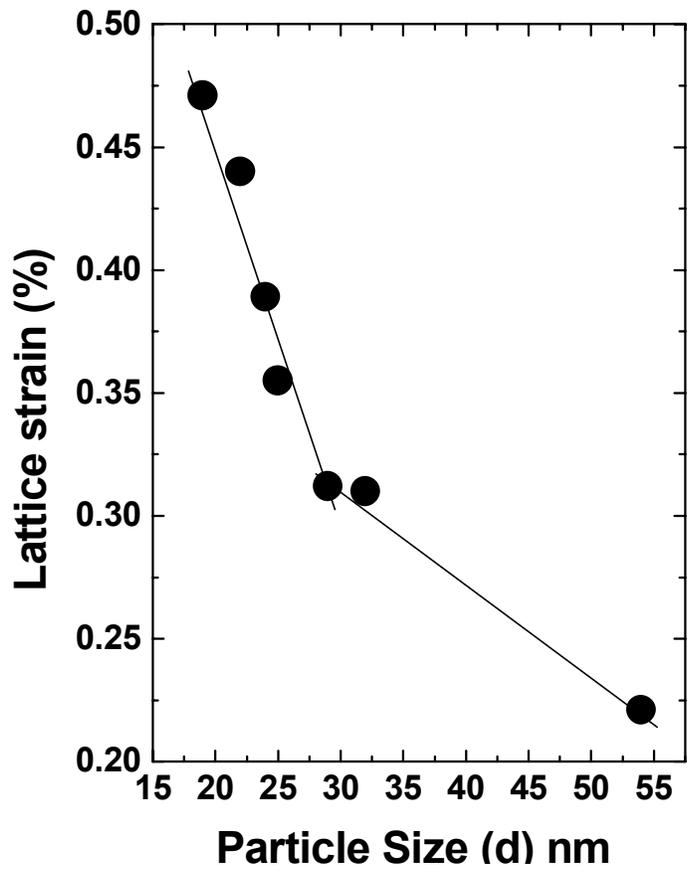

Fig. 1c

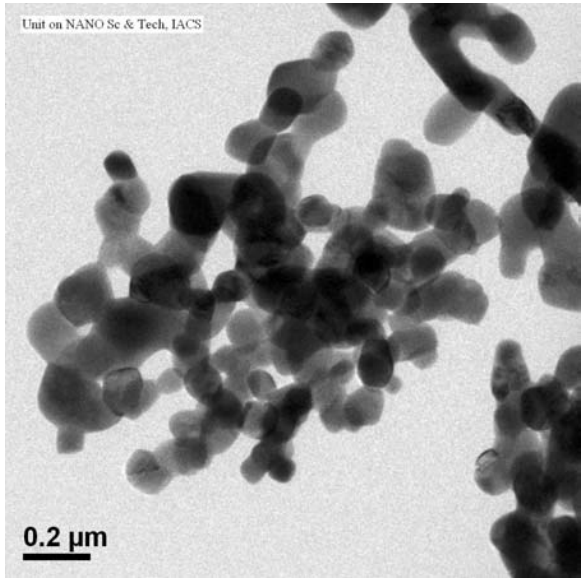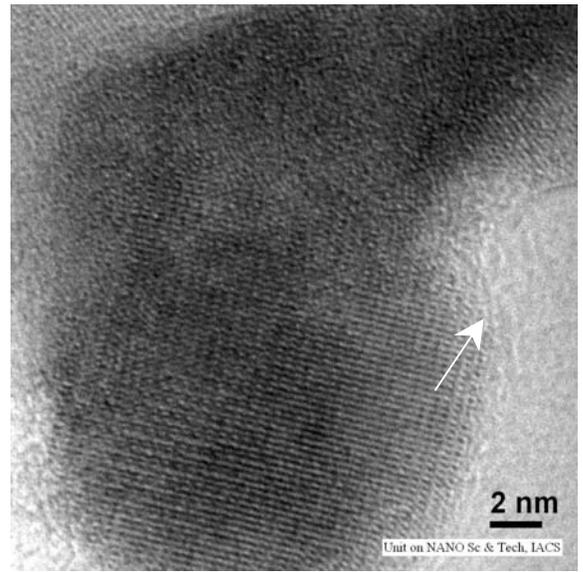
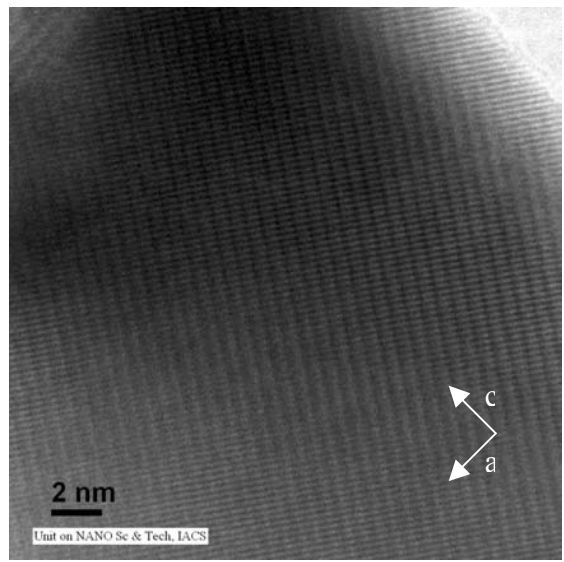

Fig. 2a

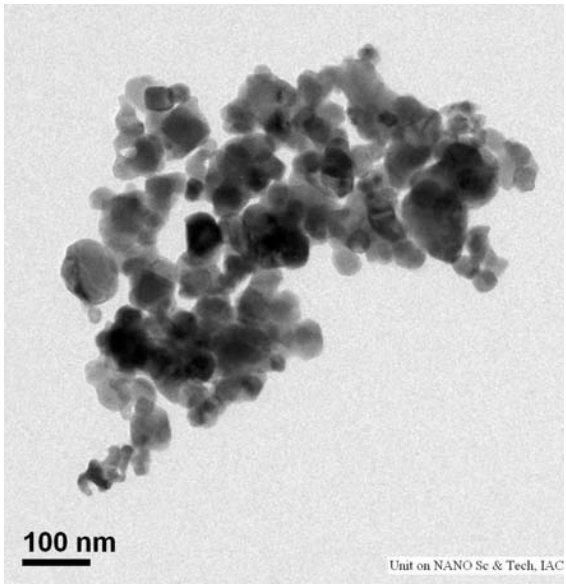
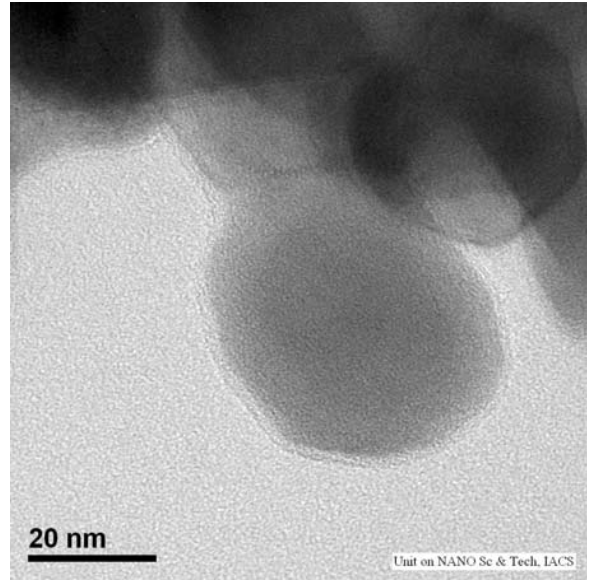
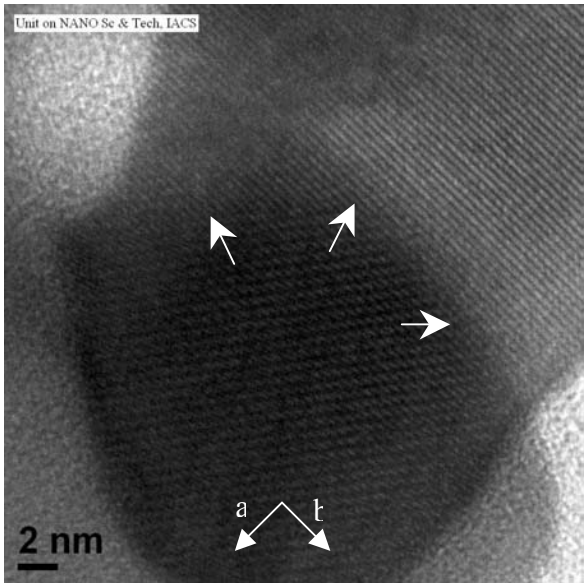

Fig. 2b

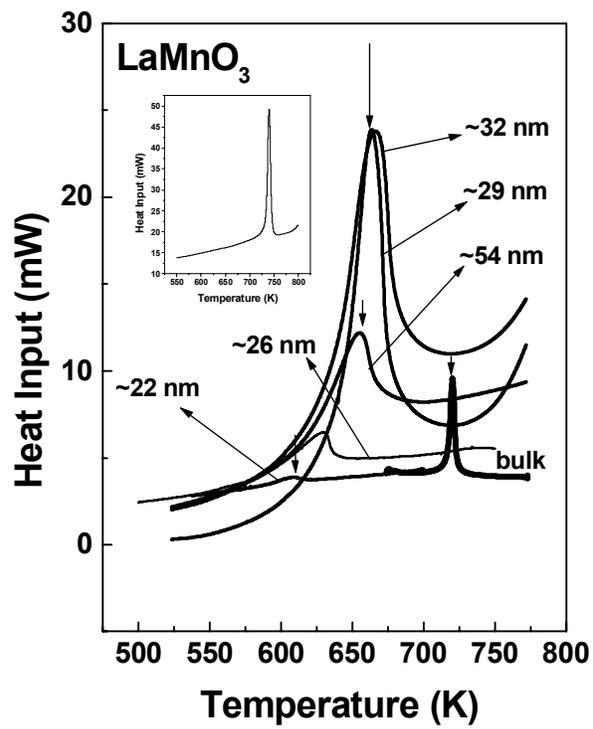

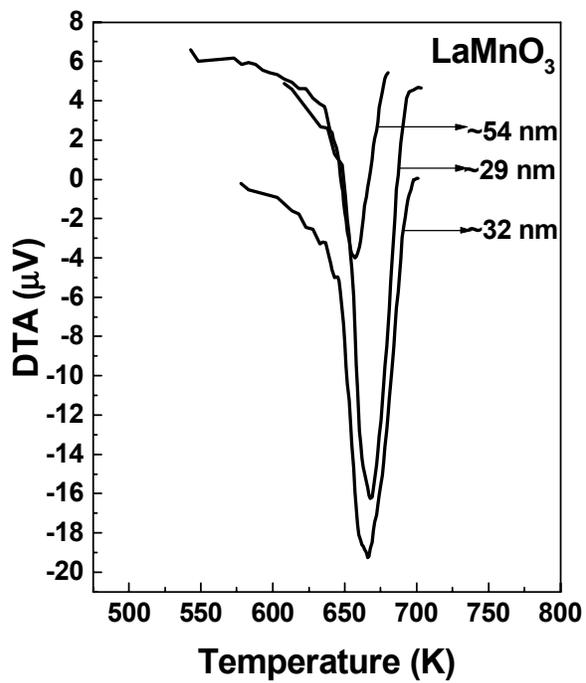

Fig. 3

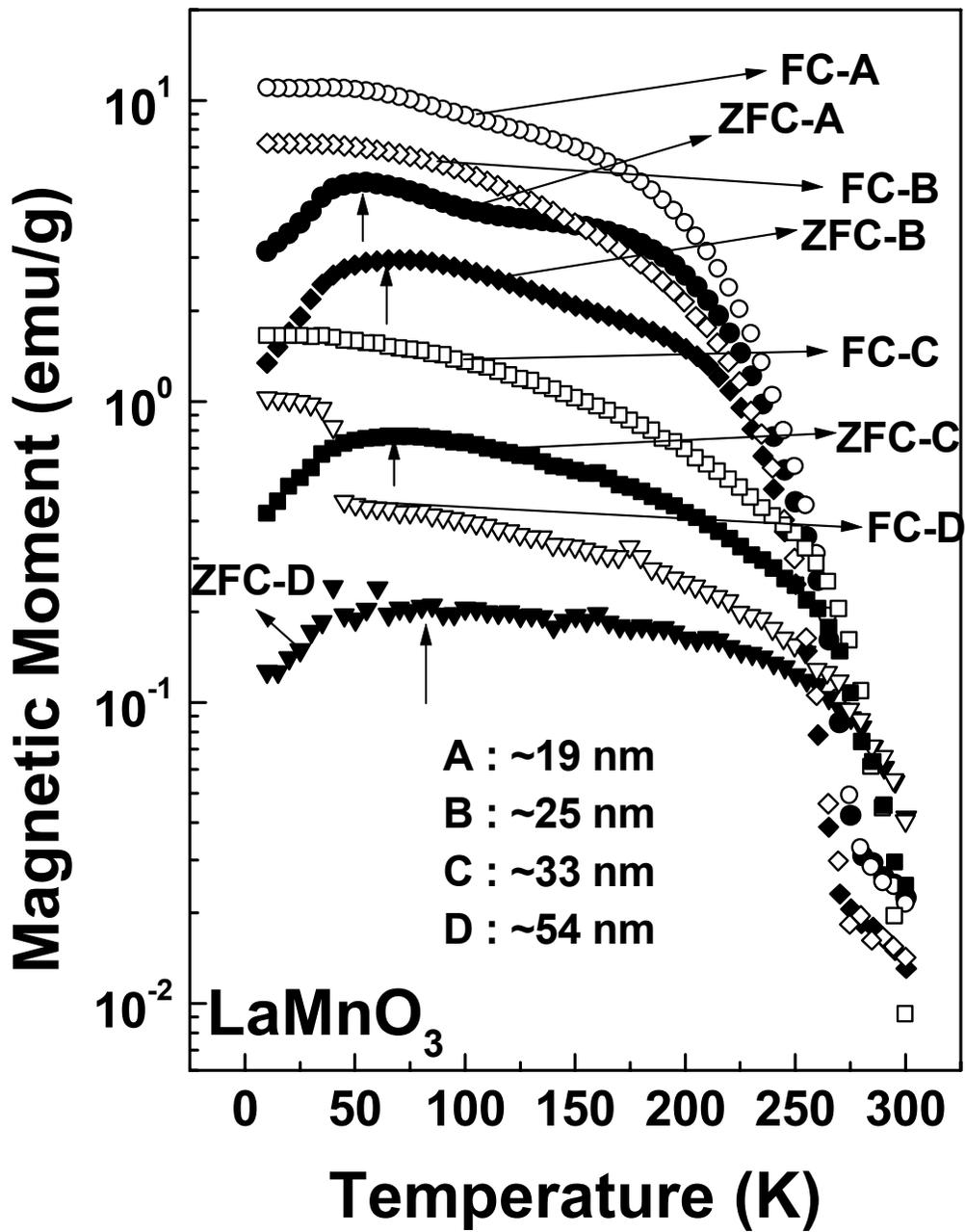

Fig. 4

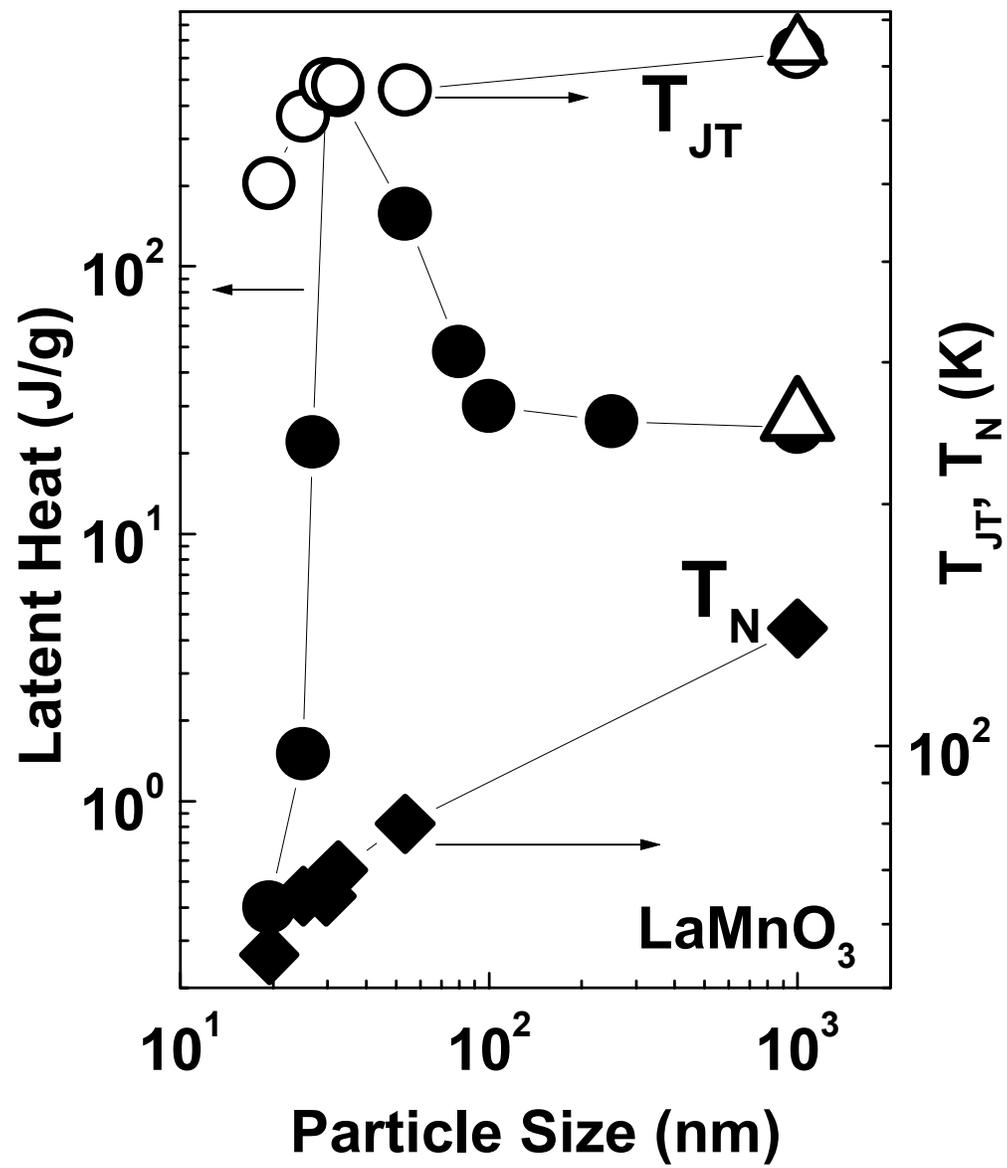

Fig. 5